\documentclass[runningheads]{llncs}
\usepackage{graphicx}
\usepackage{tikz,xcolor,hyperref}
\usepackage{comment}
\usepackage{amssymb} 
\usepackage{caption} 
\usepackage{subcaption} 
\usepackage{float}

\begin{document}
\title{Towards Knowledge Graphs Validation through Weighted Knowledge Sources}
\author{
Elwin Huaman\inst{1}\orcidID{0000-0002-2410-4977} \and
Amar Tauqeer\inst{1}\orcidID{0000-0002-3345-387X} \and
Anna Fensel\inst{1,2}\orcidID{0000-0002-1391-7104}}
\authorrunning{E. Huaman et al.}
%
\institute{Semantic Technology Institute (STI) Innsbruck, \\
Department of Computer Science, University of Innsbruck, Austria \\
\email{elwin.huaman@sti2.at, amar.tauqeer@sti2.at}\\
\and
Wageningen Data Competence Center and Consumption and Healthy Lifestyles Chair Group, Wageningen University  \& Research, The Netherlands \\
\email{anna.fensel@wur.nl}}
\maketitle              
\begin{abstract}
The performance of applications, such as personal assistants and search engines, relies on high-quality knowledge bases, a.k.a. Knowledge Graphs (KGs). To ensure their quality one important task is knowledge validation, which measures the degree to which statements or triples of KGs are semantically correct. KGs inevitably contain incorrect and incomplete statements, which may hinder their adoption in business applications as they are not trustworthy.
In this paper, we propose and implement a Validator that computes a confidence score for every triple and instance in KGs. The computed score is based on finding the same instances across different weighted knowledge sources and comparing their features.
We evaluate our approach by comparing its results against a baseline validation. Our results suggest that we can validate KGs with an f-measure of at least 75\%. Time-wise, the Validator, performed a validation of 2530 instances in 15 minutes approximately. 
Furthermore, we give insights and directions toward a better architecture to tackle KG validation.
\keywords{Knowledge graph validation \and Knowledge graph curation \and Knowledge graph assessment.}
\end{abstract}
\section{Introduction}
\label{sec:introduction}
Over the last decade, creating and especially maintaining knowledge bases have gained attention, and therefore large knowledge bases, also known as knowledge graphs (KGs)~\cite{HoganBCdMGKGNNN2021}, have been created, either automatically (e.g. NELL~\cite{CarlsonBKSHM2010}), semi-automatically (e.g. DBpedia~\cite{AuerBKLCI2007}), or through crowdsourcing (e.g. Freebase~\cite{BollackerEPST2008}). 
Today, open (e.g. Wikidata) and proprietary (e.g. Knowledge Vault) KGs provide information about entities like hotels, places, restaurants, and statements about them, e.g. address, phone number, and website.
With the increasing use of KGs in personal assistant and search engine applications, the need to ensure that statements or triples in KGs are correct arises~\cite{FenselSAHKPTUW20,HuamanKF20,Paulheim2017}. For example, Google shows the fact (\textit{Gartenhotel Maria Theresia GmbH, phone, 05223 563130}), which might be wrong because the $phone$ number of the $Gartenhotel$ is \textit{05223 56313}, moreover, there will be cases where the $phone$ number is not up-to-date or is missing~\cite{KarleFTF16}.
%

To face this challenge, we developed an approach to validate a KG against different knowledge sources. Our approach involves 
(1) mapping the different knowledge sources to a common schema (e.g. Schema.org\footnote{\url{https://schema.org/}}), 
(2) instance matching that ensures that we are comparing the same entity across the different knowledge sources, 
(3) confidence measurement, which computes a confidence score for each triple and instance in the KG, and 
(4) visualization that offers an interface to interact with. Furthermore, we describe use cases where our approach can be used.

There have been a few approaches proposed to validate KGs. In this paper, we review methods, tools, and benchmarks for knowledge validation. We found out that most of them focus on validating knowledge against the Web or Wikipedia. For example, the approaches measure the degree to which a statement (e.g. \textit{Paris is the capital of France}) is true based on the number of occurrences of the statement in sources such as Wikipedia, websites, and/or textual corpora. In addition, to the best of our knowledge, no studies have investigated how to validate KGs by collecting matched instances from other weighted structured knowledge sources.

%
%
In this paper, we propose a weighted approach that validates a KG against a set of weighted knowledge sources, which have different weight (or degree of importance) for different application scenarios. For example, users can define the degree of importance of a knowledge source according to the task at hand.
We validate a KG by finding the same instances across different knowledge sources, comparing their features, and scoring them. The score ranges from $0$ to $1$, which indicates the degree to which an instance is semantically correct for the task at hand.

This paper is structured as follows. Section~\ref{sec:literature-review} presents related state-of-the-art methods, tools, and benchmarks. Section~\ref{sec:approach} describes our validation approach. We evaluate our approach and show its results in Section~\ref{sec:evaluation}. Furthermore, in Section \ref{sec:use-cases} we list use cases where our approach may be needed. Finally, we conclude with Section~\ref{sec:conclusion}, providing some remarks and future work plans. 
\section{Literature Review}
\label{sec:literature-review}
Knowledge Validation (KV), a.k.a. fact checking, is the task of assessing how likely a given fact or statement is true or semantically correct~\cite{Gad-Elrab0UW2019b,LehmannGMN12,Rula2019}. 
There are currently several state-of-the-art methods and tools available that are suitable for KV. One of the prior works on automating this task focuses on analysing trustworthiness factors of web search results (e.g. the trustworthiness of web pages based on topic majority, which computes the number of pages related to a query)~\cite{NakamuraKJOKTOT07}. 
Another approach is proposed by Yin et al. ~\cite{YinHY2008}. Here, the authors define the trustworthiness of a website based on the confidence of facts provided by the website, for instance, they propose an algorithm called \textit{TruthFinder}. Moreover,~\cite{DongGHHLMSSZ14} present Knowledge Vault, which is a probabilistic knowledge base that combines information extraction and machine learning techniques to compute the probability that a statement is correct. The computed score is based on knowledge extracted from the Web and corroborative paths found on Freebase. However, the Web can yield noisy data and prior knowledge bases may be incomplete.
Therefore, we propose an approach that not only takes into account the user's preferences for weighting knowledge sources, but also complements the existing probabilistic approaches.
%

We surveyed methods for validating statements in KGs and we distinguish them according to the data used by them, as follows: a) \textit{internal} approaches use the knowledge graph itself as input and b) \textit{external} approaches use external data sources (e.g. DBpedia) as input. 
In the context of this paper, we only consider the approaches that use external knowledge sources for validating statements.

The external approaches use external sources like the DBpedia source to validate a statement. For instance, there are approaches that use websites information~\cite{DongGHHLMSSZ14,GerberELBUNS2015,SpeckN19}, Wikipedia pages~\cite{ErcanEH19,PadiaFF18,SyedRN2018}, DBpedia knowledge base~\cite{LiudM2015,Rula2019}, and so on.
In contrast to other approaches,~\cite{LiudM2015} present an early stage approach that uses DBpedia to find out $sameAs$ links, which are followed for retrieving evidence triples in other knowledge sources and~\cite{Rula2019} uses DBpedia to retrieve temporal constraints for a fact. However,~\cite{LiudM2015} do not provide an evaluation of the approach to be compared with our approach and~\cite{Rula2019} focus on validating dynamic data, which we do not tackle in the scope of this paper.
Furthermore, there are methods that use topic coherence \cite{AletrasS13} and information extraction \cite{SpeckN19} techniques to validate knowledge. Obviously, there is not only one approach or ideal solution to validate KGs. 
The proposed tools -- DeFacto\footnote{\url{https://github.com/DeFacto/DeFacto}}, Leopard\footnote{\url{https://github.com/dice-group/Leopard}}, FactCheck\footnote{\url{https://github.com/dice-group/FactCheck}}, and FacTify\footnote{\url{http://qweb.cs.aau.dk/factify/}}-- rely on the Web and/or external knowledge sources like Wikipedia.

The current Web-based approaches can effectively validate knowledge that is well disseminated on the Web, e.g. \textit{Albert Einstein's date of birth is March 14, 1879}. Furthermore, the confidence score is based on the number of occurrences of a statement in a corpus (e.g. Wikipedia). Unfortunately these approaches are also prone to spamming~\cite{TanAIG14}. Therefore, a new approach is necessary to further improve KG validation.
In this paper, we propose a KG validation approach, which computes a confidence score for each triple and instance of KGs. 

Furthermore, an evaluation of validation approaches is really important, therefore, we also surveyed knowledge validation benchmarks that have been proposed, however, the number of them is currently rather limited. \cite{VlachosR14} and \cite{ErcanEH19} released a benchmark consisting of triples extracted from a KG (e.g. Yago) and textual evidences retrieved from a corpus (e.g., Wikipedia). Furthermore. \cite{ThorneVCM18} released FEVER\footnote{\url{https://github.com/sheffieldnlp/fever-naacl-2018}} that is a dataset containing 185K claims about entities which were verified using Wikipedia articles. Moreover, 
FactBench\footnote{\url{https://github.com/DeFacto/FactBench}} (Fact Validation Benchmark) provides a multilingual (i.e. English, German and French) benchmark that describes several relations (e.g. Award, Birth, Death, Foundation Place) of entities.

All benchmarks mentioned above have focused mostly
on textual sources, i.e. unstructured information. Therefore, from the best of our knowledge, there is no available benchmark that can be used for validating knowledge graphs via collecting matched instances from other structured knowledge sources. 

Last but not least, the reviewed approaches are mostly focused on validating well disseminated knowledge than factual knowledge. Furthermore, benchmarks are built for validating specific tools or to be used during contests like FEVER. Another interesting observation is that Wikipedia is the most frequently used by external approaches (i.e. Wikipedia as textual corpus for finding evidences). Finally, to make future works on knowledge graph validation comparable, it would be useful to have a common selection of benchmarks.
\section{Approach}
\label{sec:approach}
\begin{figure}
    \centering
    \includegraphics[width=\textwidth]{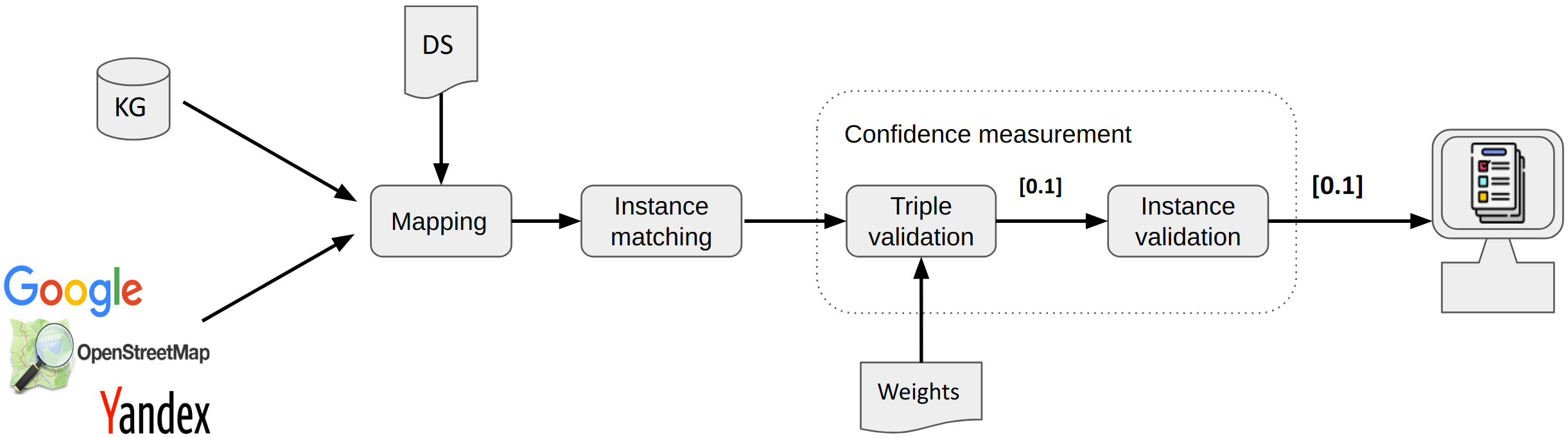}
    \caption{Knowledge graph validation process overview.}
    \label{fig:knowledge-validation-process}
\end{figure}

In this section, we present the conceptualization of our KG validation approach. 
First, we give an overview of the knowledge validation process (see Fig.~\ref{fig:knowledge-validation-process}). 
Second, we state the input needed for our approach in Section~\ref{subsec:input}.
In Section~\ref{subsec:mapping}, we describe the need for a common attribute space between knowledge sources. Then, in Section~\ref{subsec:instanceMatching}, we explain the instance matching process. Afterwards, confidence measurement of instances is detailed in Section~\ref{subsec:confidenceMeasurement}. Finally, in Section~\ref{subsec:output}, we describe the output of our implemented approach.

The input to the Validator is a KG, which can be provided via a SPARQL endpoint or an RDF dataset in Turtle\footnote{\url{https://www.w3.org/TR/turtle/}} format. 
This input KG is first mapped based on a Domain Specification\footnote{Domain Specification are design patterns for annotating data based on Schema.org. This process implies to remove types and properties from Schema.org, or add types and properties defined in an external extension of Schema.org.} (DS), which basically defines the mapping of the KG to a common format, e.g., this process may be performed by a domain expert, who defines the types and properties that are relevant to the task at hand or the user's need~\cite{SimsekAKPF2020}. A DS defines the instance type and properties values to be validated.
Internally, the Validator is configured to retrieve data from external sources, which are also mapped to the common format. 
After the mapping process has been done, the instance matching is used to find the same instances across the KG and the external sources. Then, the confidence measurement process is triggered and the features of same instances are compared with each other. For example, we compare the name value of an instance of the KG against the name value of the same instance in an external source. We repeat this process for every triple of an instance and we compute a triple confidence score, the triple confidence scores are later added to an aggregated confidence score for the instance. The computed scores are normalized according to the weights given to each knowledge source. We consider the quality of the external sources subjective, therefore, we provide a graphical user interface that allows users to weight each knowledge source.

\subsection{Input} 
\label{subsec:input}
At first step, a user is required to provide a KG to be validated.
For this, the user has two options, a) to provide a SPARQL endpoint where to fetch the data from or b) to load a dataset in a Turtle format. 
Moreover, the user is required to select, from a list of DSs, a DS that defines an instance type (e.g., Hotel, Person) and their corresponding properties (e.g., name, address). 
Internally, the Validator has been set up to fetch data from different external sources (e.g. Wikidata, DBpedia), which were selected based on their domain coverage for the task at hand and their widely use~\cite{FarberBMR18}.

\subsection{Mapping} 
\label{subsec:mapping}
Based on the DS defined in the input, the validator maps the input KG and the external sources to a common format, e.g., a telephone number of a hotel can be stored with different property names across the knowledge sources: \textit{phone}, \textit{telephone}, or \textit{phone\_number}. The validator provides a basic mapping feature to map the input KG and external data sources to a common attribute space. 
This step is not trivial. There is a huge number of knowledge sources and their schemas might be constantly changing~\cite{DongS2015}. As a result,
schema alignment\footnote{Schema alignment is the task of determining the correspondences between various schemas.} is one of the major bottlenecks in the mapping process. Therefore, new methods and frameworks to tackle the schema heterogeneity are needed.

\subsection{Instance Matching} 
\label{subsec:instanceMatching}
So far, we mapped knowledge sources to a common attribute space. However, a major challenge is to match instances across these knowledge sources. For that, the Validator requests to define at least two or more properties (e.g., \textit{name} and \textit{geo coordinates}) that are to be used for the instance matching process, which is constrained to strict matches on the defined property values. The resulting matched instance is returned to the Validator and processed to measure its confidence.
\subsection{Confidence Measurement} 
\label{subsec:confidenceMeasurement}
Computing a confidence value can get complicated as the number of instances and their features can get out of hand quickly. Therefore, a means to automatically validate KGs is desirable. To compute a confidence value for an instance, the confidence value for each of its triples has to be evaluated first.
\subsubsection{Triple validation}
\label{subsubsec:triple-validation}
calculates a confidence score of whether a property value on various external sources matches the property value in the user's KG. For example, the user's KG contains the \textit{Hotel Alpenhof} instance and statements about it; \textit{Hotel Alpenhof's phone is +4352878550} and \textit{Hotel Alpenhof's address is Hintertux 750}. Furthermore, there are other sources, like Google Places, that also contain the \textit{Hotel Alpenhof} instance and  assertions about it. 

The confidence score of \textit{(Hotel Alpenhof, phone, +4352878550)} triple is computed by comparing the phone property value \textit{+4352878550} against the same property value of the same instance in Google Places. For that, syntactic similarity matching of the attribute values is used. Then the phone property value is compared against a second knowledge source, and so on. Every similarity comparison returns a confidence value that later is added to an aggregated score for the triple.

We define a set of knowledge sources as $S$, \(S = \{s_1, \dots ,s_m\}\),
\(s_{i} \in S\) with \(1 \leq i \leq m\). 
The user's KG $g$ consists of a set of instances that are to be validated against the set of knowledge sources $S$. A knowledge source \(s_i\) consists of a set of instances \(E = \{e_1, \dots , e_n\}\), \(e_{j} \in E\) with \(1 \leq j \leq n\) and 
an instance $e_j$ consists of a set of attribute values \(P = \{p_1, \dots , p_M\}\), \(p_{k} \in P\) for \(1 \leq k \leq M\).

Furthermore, $sim$ is a similarity function used to compare attribute pair $k$ for two instances. We compute the similarity of an attribute value of two instances $a$, $b$. Where $a$ represents an instance in the user's KG $g$, denoted $g(a)$, and $b$ represents an instance in the knowledge source $s_i$, denoted $s_i(b)$.
\begin{equation}
    triple_{confidence}(a_{p_k},S,sim) = \sum_{i=1}^{m}sim(g(a_{p_k}), s_{i}(b_{p_k}))
\end{equation}

Next, users have to set an external weight for each knowledge source \(s_{i}\), \(W = \{\omega_1, \dots , \omega_m\}\) is a set of weights over the knowledge sources, such as \(\omega_{i}\) defines a weight of importance for \(s_{i}\), \(0 \leq i \leq m\), \(\omega_{i} \in W\) with \(\omega_{i} \in [0,1]\) where \(0\) is the minimum degree of importance and a value of \(1\) is the maximum degree. 
For the sum of weights \(w_{sum} = \sum_{i=1}^{m}\omega_i = 1\) has to hold.  
We compute the weighted triple confidence as follows:

\begin{equation}
    triple_{confidence}(a_{p_k},S,sim,W) = \frac{1}{w_{sum}}\sum_{i=1}^{m}sim(g(a_{p_k}), s_{i}(b_{p_k}))w_i
\end{equation}

The weighted approach\footnote{To define weights, a proper quality analysis of the knowledge sources must be carried out~\cite{FarberBMR18}. It may assist users in defining degrees of importance for each knowledge source.} aims to model the different degrees of importance of different knowledge sources. None of the parameters can be taken out of their context, thus a default weight has to be given whenever the user does not set weights for an external source. The Validator assigns an equivalent weight for each source: \( \omega_{i} = \frac{1}{m} \).
\subsubsection{Instance validation}
\label{subsubsec:instance-validation}
computes the aggregated score from the attribute space of an instance. Given an instance $a$ that consists of a set of attribute values \(P = \{p_1, \dots , p_M\}\), \(p_{k} \in P\) for \(1 \leq k \leq M\):
\begin{equation}
    instance_{confidence}(a_{p_k},S,sim,W) = \frac{1}{M}\sum_{k=1}^{M}triple_{confidence}(a_{p_k},S,sim,W)
\end{equation}
The instance confidence measures the degree to which an instance is correct based on the triple confidence of each of its attributes. The instance confidence score is compared against a threshold\footnote{The default threshold is defined to 0.5} $t \in [0,1]$. If $instance_{confidence} > t$ indicates its degree of correctness.
\begin{figure}
	\includegraphics[width=\textwidth]{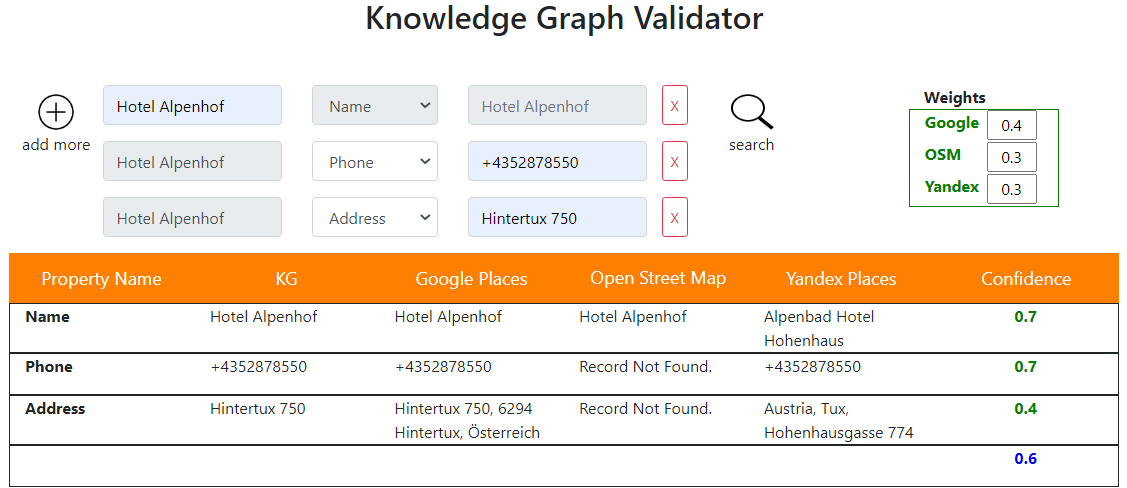}
	\caption{Screenshot of the Validator Web interface.}
	\vspace{-3em}
	\label{fig:validator-screenshot}
\end{figure}
\subsection{Output} 
\label{subsec:output}
The computed scores for triples and instances are shown in a graphical user interface, see Fig.~\ref{fig:validator-screenshot}. The interface provides many features: it allows users to select multiple properties (e.g. address, name) to be validated, users can assign weights to external sources, it shows instance information from user's KG and external sources. For example, the Validator shows information of the \textit{Hotel Alpenhof} instance from all sources. It also shows the triple confidence score for each triple, e.g. the triple confidence for the address property is $0.4$, because the address value is confirmed only by Google Places.
\subsubsection{Tools \& Technologies.}
\label{subsubsec:tools-technology}
We implemented our approach in the Validator tool\footnote{\url{https://github.com/AmarTauqeer/graph-validation}}, which has been implemented in JavaScript\footnote{\url{https://developer.mozilla.org/en-US/docs/Web/JavaScript}} for retrieving data remotely, and Bootstrap\footnote{\url{https://getbootstrap.com/}} for the user interface.
%
%
\section{Evaluation}
\label{sec:evaluation}
This section describes the evaluation of our approach. The aim of the experiments is to show a qualitative and quantitative analysis of our approach. The setup used for the evaluation is described in Table~\ref{tab:computer-setup}.

\begin{table}[h!]
    \centering
    \begin{tabular}{|c|c|c|}
        \hline
        CPU & RAM & OS \\
        \hline\hline
        AMD Ryzen 7 pro 4750u (16 Cores) & 32GB & Ubuntu 20.04.2 LTS 64-bit \\
        \hline
    \end{tabular}
    \caption{Evaluation setup}
    \label{tab:computer-setup}
\end{table}

In Section~\ref{subsubsec:qualitative-eval}, we compare the Validator's validation result against a baseline. Next, we look into the scalability of the Validator in Section~\ref{subsubsec:scalability-eval}.
\subsection{Qualitative evaluation}
\label{subsubsec:qualitative-eval}
The qualitative evaluation measures the effectiveness of the Validator based on a baseline validation. To do so, first, we describe a dataset to be used on the quality evaluation of our approach, later on we define a setup for the Validator and execute it. Then, we stablish a baseline to compare the result of the Validator. 

\subsubsection{Hotel dataset.}
\label{subsubsec:quality-eval-dataset}
It was fetched from the Tirol Knowledge Graph\footnote{\url{https://graphdb.sti2.at/sparql}} (TKG), which contains \(\sim15\) Billion statements about hotels, places, and more, of the Tirol region. The data inside the TKG are static (e.g name, phone number) and dynamic (e.g. availability of rooms, prices) and are based on Schema.org annotations, which are collected from different sources such as destination management organizations and geographical information systems. We have created a benchmark dataset of 50 hotel instances\footnote{\url{https://github.com/AmarTauqeer/graph-validation/tree/master/data}} fetched from the TKG. We randomly selected 50 hotel instances in order to be able to perform a manual validation of their correctness and establish a baseline. The process of creating the Hotel dataset involved manual checking of the correctness of all instances and their attribute values.

\subsubsection{Setup and Execution.}
\label{subsubsec:quality-eval-setup}
First, we set up the Hotel dataset on the Validator.
Second, we defined external sources, namely: Google Places\footnote{\url{https://developers.google.com/maps/documentation/places/}}, OpenStreetMap (OSM)\footnote{\url{https://www.openstreetmap.org/}}, and Yandex Places\footnote{\url{https://yandex.com/dev/maps/}}. 
Third, we defined the \textit{Hotel} type and \textit{address}, \textit{name}, and \textit{phone} properties that are used for mapping place instances from external sources. Then, for the instance matching process, we set up the \textit{name} and \textit{geo-coordinates} values to search for places within a specified area. We use the built-in feature provided by the external sources (e.g. Nearby Search for Google places) to search for an instance with the same name within a specific area. Furthermore, weights for the external sources are equally distributed. Finally, we run the validation task.

\subsubsection{Baseline.}
\label{subsubsec:quality-eval-baseline}
\begin{figure}
    \centering
    \includegraphics[width=0.9\textwidth]{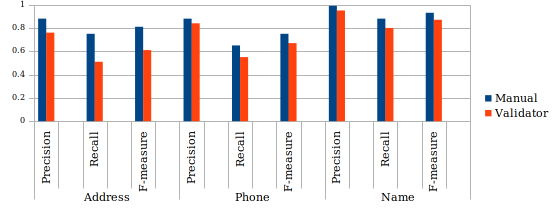}
    \caption{Comparison of precision, recall, and f-measure scores over the manual and semi-automatic validation.}
    \label{fig:quality-eval-result}
\end{figure}

In order to evaluate the results of the Validator, a baseline must be established. Given that no prior validation tool addresses exactly the task at hand, we made a manual validation of the Hotel dataset. We computed the precision, recall, and f-measure that a manual validation would achieve (See Fig.~\ref{fig:quality-eval-result}). During this evaluation, the 50 hotel instances are manually searched and compared to the results coming from each of the external knowledge sources: Google Places, OSM, and Yandex Places. The compared attributes are the \textit{address, name}, and \textit{phone}. 
\subsubsection{Result.}
\label{subsubsec:quality-eval-result}
We analyse the result of running the Validator on the Hotel dataset. These results are shown in Fig.~\ref{fig:quality-eval-result}. On one hand, it shows that the Validator performs almost equally similar as the manual evaluation when it comes to \textit{name} and \textit{phone} properties, on the other hand, the Validator does not perform well on the validation of the \textit{address} property. Moreover, the results suggest that we can validate hotel instances with an f-measure of at least 75\% on \textit{address}, \textit{name}, and \textit{phone} properties. 
To interpret the results of our validation run, we choose precision, recall, and f-measure. Given the results of the Validator run, every validated triple result was classified as True Positive, False Positive, True Negative, or False Negative based on the baseline results.

\subsection{Scalability evaluation}
\label{subsubsec:scalability-eval}
Another challenge of a validation framework is the scalability. In this section, we describe our evaluation approach in terms of scalability of our approach.
\subsubsection{Pantheon dataset.}
\label{subsubsec:quantity-eval-dataset}
It contains manually validated data with 11341 famous biographies~\cite{Yu2016}. Pantheon describes information like name, year of birth, place of birth, occupation, and many more. We have selected politician domain and created a dataset of 2530 politician instances. We selected the politician domain because it has the highest number of instances in the Pantheon dataset. Furthermore, we had to convert the Pantheon dataset to Turtle format, for that we used Tarql\footnote{\url{https://tarql.github.io/}} tool. Last but not least, we selected the politician domain in order to prove the general applicability of our approach in different domains (e.g., Hotel, Person). 
\subsubsection{Setup and Execution.}
\label{subsubsec:quantity-eval-setup}
The setup for validating datasets from different domains changes slightly, for example, defining the external sources where to fetch the data from. 
First, we set up the Pantheon dataset on the Validator. Then, we defined Wikidata and DBpedia as external sources and we distributed equivalent weights for them. Moreover, we defined the \textit{person} type and \textit{name} and \textit{year of birth} properties for mapping politicians from the external sources. Moreover, we set up the \textit{name} and \textit{year of birth} for the instance matching process. Finally, we execute the validation task. 
%
\begin{figure}
    \centering
    \includegraphics[width=0.5\textwidth]{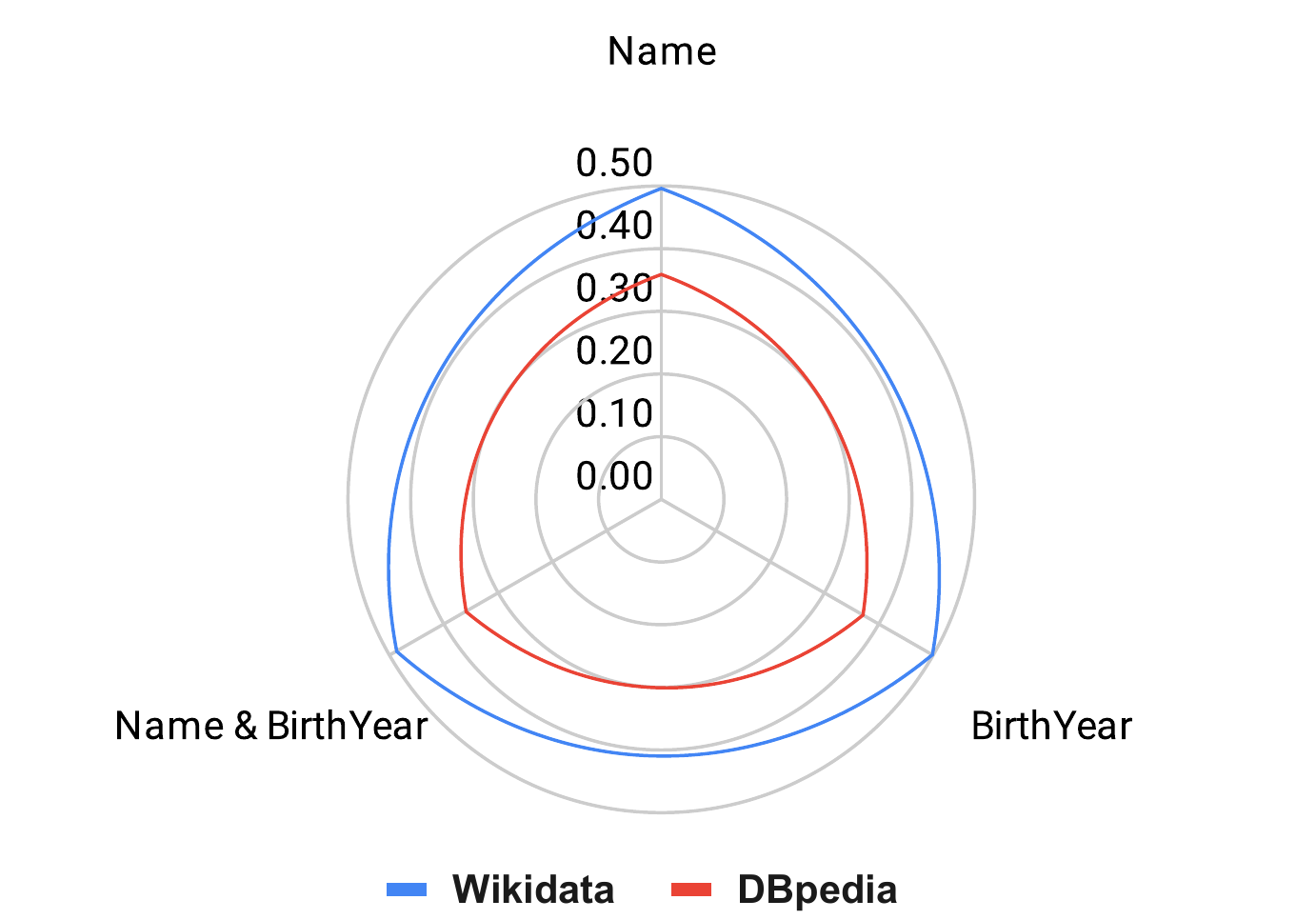}
    \caption{The recall score results of the validation of politician instances.}
    \label{fig:pantheon-result}
\end{figure}
%
\subsubsection{Result.}
\label{subsubsec:quantity-eval-result}
We validated 2530 politician instances by using the Validator, which compares and computes a confidence score for each triple and instance. To execute this task the Validator required \(\sim\)15 minutes approximately on a CPU described on Table~\ref{tab:computer-setup}. Results are presented in Fig.~\ref{fig:pantheon-result}. On one hand, it shows that Wikidata outperforms DBpedia on validated properties, on the other hand, it shows lower recall scores, by the Validator, on both sources, e.g. the overall recall scores are $0.36\%$ (DBpedia) and $0.49\%$ (Wikidata).

Furthermore, the Validator gets lower recall on DBpedia and Wikidata sources due to two reasons. First, DBpedia contains the validated politician instances, however many of them are classified in DBpedia as \textit{agent} type and not as politician (e.g., \textit{Juan Carlos I}\footnote{\url{https://dbpedia.org/page/Juan_Carlos_I}}). Second, the Wikidata query service raised timeout errors when querying data, so we decided to fetch the maximum allowed number of politician instances from Wikidata and stored them locally. We fetched 45000 out of 670810 politicians.
\section{Use Cases}
\label{sec:use-cases}
Our approach, as described in Section \ref{sec:approach}, aims to validate KGs by finding the same instances across different knowledge sources and comparing their features. Later on, based on the compared features our approach computes a confidence score for each triple and instance, the confidence score ranges from $0$ to $1$ and indicates the degree to which an instance is correct. Our approach may be used in a variety of use cases, we list some of the cases where the approach can be used:

\begin{itemize}
    \item To validate the semantic correctness of a triple, e.g., to validate if the phone number of a hotel is the correct based on different sources.
    \item To link instances between knowledge sources, e.g. linking an instance of the user's KG with the matched instance in Wikidata.
    \item To find out incorrect data on different knowledge sources. For instance, suppose that the owner of a hotel wants to validate whether the information of his or her hotel provided by an external source are up-to-date.
    \item To validate static data, for example, to check whether the addresses of hotels are still valid given a period of time.
\end{itemize}

There are more possible use cases where our validation approach is applicable. Here, we presented some of them to give an idea about how useful and necessary is to have a validated KG (i.e. a correct and reliable KG).
\section{Conclusion and Future Work}
\label{sec:conclusion}
In this paper, we presented the conceptualization of a new KG validation approach and a first prototypical implementation thereof. Our approach measures the degree to which every instance in a KG is semantically correct. It evaluates the correctness of instances based on external sources. Experiments were conducted on two datasets. The results confirm its effectiveness and are promising great potential. In future work, we will improve our approach and overcome its limitations. Here, we give a short overview of them:
\begin{itemize}
    \item \textbf{Assessment} of knowledge sources. Finding the most suitable knowledge source for validating a KG is challenging~\cite{FarberBMR18}. Therefore, it is desirable to implement a quality assessment mechanism for assessing external sources. It may assist users in defining degrees of importance for each knowledge source.
    \item \textbf{Automation} of the setting process. It is desirable to allow users to create a semi-automatic \textbf{mapping} (or schema alignment~\cite{DongS2015}) between their KG and the external sources, e.g. the heterogeneous scheme of OSM has caused low performance of the Validator (see Section~\ref{subsubsec:quality-eval-result}).
    \item \textbf{Cost-sensitive methods}. The current version of the Validator relies on proprietary services like Google, which can lead to high costs when validating large KGs. Therefore, it is important to evaluate the cost-effectiveness of knowledge sources.
    \item \textbf{Dynamic data} is fast-changing data that also needs to be validated, e.g. the price of a hotel room. The scope of this paper only comprises the validation of static data. 
    \item \textbf{Scalability} is a critical point when we want to validate KGs. KGs are very large semantic networks that can contain billions of statements.
\end{itemize}
Above, we pointed out some future research directions and improvements that one can implement on the development of future validation tools.\\

\textbf{Acknowledgments.} This work has been partially funded by the project WordLiftNG within the Eureka, Eurostars Programme of the European Union (grant agreement number 877857 with the Austrian Research Promotion Agency (FFG)) and the industrial research project MindLab\footnote{\url{https://mindlab.ai/}}. We would like to thank Prof. Dr. Dieter Fensel for his insightful comments regarding the definition of the overall validation approach.
\bibliographystyle{splncs04}
\bibliography{bibliography}
\end{document}